\documentclass[12pt]{iopart}
\usepackage{graphicx}

\newcommand{\eqd}{\stackrel{\triangle}{=}}

\newcommand {\exe} {\stackrel{\cdot} {=}}

\newcommand {\ba} {\mbox{\boldmath $a$}}

\newcommand{\hsig}{{\hat{\sigma}}}

\newcommand {\bx} {\mbox{\boldmath $x$}}

\newcommand {\bsigma} {\mbox{\boldmath $\sigma$}}
\newcommand {\balpha} {\mbox{\boldmath $\alpha$}}
\newcommand {\btau} {\mbox{\boldmath $\tau$}}

\newcommand{\calI}{{\cal I}}

\newcommand{\calN}{{\cal N}}

\newcommand{\calS}{{\cal S}}

\begin{document}

\title[Subset--sum data compression]{Subset--sum
phase transitions and data compression}

\author{Neri Merhav}

\address{Department of Electrical Engineering, Technion, Haifa 32000,
Israel.\\ E--mail: merhav@ee.technion.ac.il}

\begin{abstract}
We propose a rigorous analysis 
approach for the subset sum problem in the context of
lossless data compression, where the phase transition of the subset sum
problem is directly related to the passage between ambiguous and non--ambiguous
decompression, for a compression scheme that is based on
specifying the sequence composition. The proposed analysis lends itself to
straightforward extensions in several directions of interest, including
non--binary alphabets, incorporation of side information at the decoder
(Slepian--Wolf coding), and
coding schemes based on multiple subset sums. It is also demonstrated that the
proposed technique can be used to analyze the critical behavior in a more
involved situation where the sequence composition is not specified by the
encoder.
\end{abstract}

\maketitle

\section{Introduction}

We consider a lossless data compression scheme that builds upon
the the {\it number partitioning} problem and the closely related
problem of {\it subset sums}: Given a set of integers, $a_1,a_2,\ldots,a_N$,
$a_i\in\{1,2,\ldots,L\}$, $i=1,2,\ldots,N$, the number partitioning problem
is the problem of finding a subset $\calS\subseteq\{1,2,\ldots,N\}$, such that
the sums of $\{a_i\}$ over $\calS$ would be as balanced as possible with the
sum over the remaining $\{a_i\}$. More precisely, the goal is to find a subset
$\calS$ such that
$|\sum_{i\in\calS}a_i-\sum_{i\in\calS^c}a_i|$ would be minimum, or
equivalently, to find a binary vector 
$\bsigma=(\sigma_1,\ldots,\sigma_N)\in\{-1,+1\}^N$
such that $|\sum_{i=1}^Na_i\sigma_i|$ would be minimum.
Perfect partitioning means that this expression is exactly equal to zero.
The problem of finding an optimum partition is NP--complete
\cite {GJ79}, \cite{Papadimitriou94} and it has a
fairly long history
(see, e.g., \cite[Section 9.2]{Nishimori01}, \cite[Chapter 7]{MM09} and many
references therein).
For the case where $\{a_i\}$ are drawn independently at random, some
rigorous results have been obtained using methods of statistical mechanics,
see, e.g, \cite{BCP01}, \cite{BCMP03}, \cite{FF98}, \cite{Mertens01}.
It has been
shown (see also \cite{MM09}, \cite{Nishimori01}, \cite{STN01})
that for a randomly selected vector $(a_1,\ldots,a_N)$, and for
$L=2^{NR}$ ($R > 0$, constant), there is a phase transition at $R=1$. For $R <
1$, there are exponentially many solutions ($\sigma$--vectors) to the
number partitioning problem. More precisely,
there are exponentially about $2^{N(1-R)}$
many solutions on the average. 
However, for $R > 1$, the probability that there exists even one
solution decays exponentially.

In the related problem of subset sums,
the scope is extended to the evaluation of the total number $\Omega(E)$ of binary vectors
$\{\bsigma\}$ such that $\sum_{i=1}^Na_i\sigma_i=E$, for any
given value of $E$ in the appropriate range, not only $E=0$. 
Sasamoto \cite{Sasamoto03} proposed 
a data compression scheme based on subset sums in its constrained version,
that is, the one where binary vectors are sought only among those which have a given
{\it composition},
namely, given numbers $N_+=Np$ ($0\le p\le 1$) and $N_-=Nq$ ($q=1-p$) of occurrences of $\sigma_i=+1$ and 
$\sigma_i=-1$, respectively, or
equivalently, a given value of
$M(\bsigma)=\sum_{i=1}^N\sigma_i=N(p-q)$.\footnote{By contrast, in the
unconstrained version, solutions are sought across all binary strings of
length $N$.}
In particular,
in view of the above described results concerning phase transitions,
Sasamoto's insight was that for $R$ above a certain threshold, the mapping between
the set of binary
vectors $\{\bsigma\}$ of a given composition
to the sums $E(\bsigma)=\sum_{i=1}^Na_i\sigma_i$ must be essentially
one--to-one for a typical realization of $(a_1,\ldots,a_N)$. 
This has lead him to propose a lossless data compression scheme that is based
on encoding a binary string $\bsigma$, with a composition of $N_+=Np$ and
$N_-=Nq$, using a binary representation of $E(\bsigma)$ plus a relatively
small overhead (of $\log(N+1)$ bits) for specifying the composition of $\bsigma$, or
equivalently, the value of $M(\bsigma)=\sum_{i=1}^N\sigma_i$.
Sasamoto argued that the threshold of reliable decoding occurs at
$R=h(p)$, where 
\begin{equation}
h(p)=-p\log p-(1-p)\log(1-p) 
\end{equation}
is entropy of the binary 
information source that emits sequences with the aforementioned composition
(within some small tolerance) with high probability,
and so by taking $R=h(p)+\epsilon$
($\epsilon > 0$, arbitrarily small)
and using the fact that the range of possible values of 
$E(\bsigma)$ does not exceed $N\cdot L$,
one may encode $E(\bsigma)$ using $\log(N\cdot
L)=N[h(p)+\epsilon]+\log N$ bits, and thereby essentially achieve the entropy
of the information source.
While this coding scheme is not very attractive from the practical
point of view, the interesting point here is the relationship between the phase
transition of the subset problem and the abrupt passage between ambiguous and
non--ambiguous decoding as $R$ crosses the entropy, in agreement with Shannon's
fundamental coding theorems \cite{CT06}.

Sasamoto's approach was to analyze the number $\Omega(E,M)$
of configurations $\{\bsigma\}$ with $E(\bsigma)=E$ and $M(\bsigma)=M$,
where $E$ and $M$ are the values pertaining to the source sequence
$\hat{\bsigma}=(\hat{\sigma}_1,\ldots,\hat{\sigma}_N)$ that was actually
compressed. He argued that 
for a typical realization of $\{a_i\}$, the behavior is
as follows: For $R < h(p)$, $\Omega(E,M)$
is exponentially large and so the decoding of $\hat{\bsigma}$, based on
$E$ and $M$, is ambiguous, but for $R > h(p)$, the expectation of $\Omega(E,M)$ is exponentially
small, and so the decoding is reliable with high probability.
In order to assess the number of solutions to the two simultaneous equations
$E(\bsigma)=E$ and $M(\bsigma)=M$, he applied 
the saddle point method (see also
\cite{Nishimori01}, \cite{STN01}). In particular,
he first defined a partition function of
a Hamiltonian defined by a linear combination of 
$E(\bsigma)$ and $M(\bsigma)$, and then 
used the integral representation of the inverse transform of this
partition function, that yields
$\Omega(E,M)$. This integral in turn was approximated using 
the saddle point method. 

The analysis in \cite{Sasamoto03}, which relies on 
the analysis in \cite{STN01}, raises two technical concerns, however.
The first is about the validity of the saddle point method in
this situation: While the saddle
point method is perfectly rigorous under the asymptotic regime where
$N\to\infty$ while $L$ is kept fixed, its validity becomes rather
questionable\footnote{In \cite{STN01} there are detailed discussions
about the validity of the saddle point method when
$L$ is a function of $N$ (see the ending paragraph of Section 3 on page
9559 and pp.\ 9563--9564 therein).}
in a regime where $L$ grows with $N$, especially when
the growth rate of $L$ is as fast as exponential. 
The authors of \cite{STN01} realize that the resulting approximation
is definitely not valid when $R > 1$, which yields $\Omega(E,M) < 1$. The
point, however, is that it is not quite clear whether this approximation is reliable
even when $R < 1$. The fact that the resulting approximation below $R=1$ does not lead to
an obvious absurd is not enough to guarantee that the approximation is reliable.

The second concern is
that there is a difference between calculating the expectation of
$\Omega(E,M)$ when $E$ and $M$ are fixed and deterministic, and
calculating the expectation of $\Omega(E,M)$ when
$M=M(\hat{\bsigma})$ and 
$E=E(\hat{\bsigma})=\sum_ia_i\hat{\sigma}_i$,
because the latter is a {\it random variable}.
The former quantity is what Sasamoto calculated and the latter is 
actually the relevant quantity
for analyzing the data compression scheme.
When computing the expectation of $\Omega(E(\hat{\bsigma}),M(\hat{\bsigma}))$,
the randomness of $E(\hat{\bsigma})$ is induced by
the same set of random variables $\{a_i\}$ that generate also the values of
$E(\bsigma)$ pertaining to all other
binary vectors $\{\bsigma\}$. In other words, in this calculation
both the function $\Omega(\cdot,\cdot)$ and its first
argument $E(\hat{\bsigma})$ 
fluctuate together, depending on
$\{a_i\}$. Indeed, for one thing, $\Omega(E(\hat{\bsigma}),M(\hat{\bsigma}))$ 
(and hence also its expectation) must always be at least as
large as unity (by construction), whereas the expectation of $\Omega(E,M)$ for
fixed $E$ and $M$ is shown in \cite{Sasamoto03} 
to decay exponentially to zero for $R$ above
the threshold. This is clearly a contradiction. 

In this work, we first propose a rigorous approach to evaluate
the expectation of $\Omega(E(\hat{\bsigma},M(\hat{\bsigma}))$, which
is valid for every $R\ge 0$.
Our starting point is (or can be interpreted as)
essentially the same inverse transform integral
of the above--mentioned partition function (but with a slight modification
to account for the above discussed replacement of a fixed 
$E$ by $E(\hat{\bsigma})$). 
However, unlike in \cite{Sasamoto03} and \cite{STN01}, we avoid the use of the saddle point 
method in the evaluation of this integral
and we propose a more refined analysis instead. The final result of this
analysis is similar to that of \cite{Sasamoto03}
for $R$ below the threshold, but it is not quite identical above the threshold:
We show that when the relative frequencies of $+1$ and $-1$ in
$\hat{\bsigma}$ are $p$ and $q$, respectively, and $L=2^{NR}$, 
\begin{equation}
\left<\Omega(E(\hat{\bsigma}),M(\hat{\bsigma}))\right> \exe 1+2^{N[h(p)-R]},
\end{equation}
where $\left<\cdot\right>$ denotes expectation
w.r.t.\ the randomness of $\{a_i\}$ and
$\exe$ means equality in
the exponential order sense ($a_N\exe b_N$ means that
$\frac{1}{N}\ln\frac{a_N}{b_N}\to 0$ as $N\to \infty$).
Thus, indeed there is a phase transition at
$R=h(p)$: For $R < h(p)$,
there are exponentially many source vectors that are mapped to the same value
of $E(\hat{\bsigma})$ on the average,
but for $R > h(p)$ the expected number of
additional source vectors (other than $\hat{\bsigma}$) vanishes.
Since $\Omega(E(\hat{\bsigma}),M(\hat{\bsigma}))$ is an integer--valued random variable, this
also means
(by the Chebychev inequality)
that $\mbox{Pr}\{\Omega(E(\hat{\bsigma}),M(\hat{\bsigma}))>1\}$ also vanishes
for $R> h(p)$.

While the final conclusions
of our analysis are 
essentially the same as in \cite{Sasamoto03} (for $R < h(p)$), 
the message in this paper
is three--fold: The first message is that it is not necessary to resort to the
saddle point method in this case and it is possible to make the analysis
rigorous as we show.
The second message is that our analysis
extends easily to more general situations, like larger source alphabets,
availability of side
information at the decoder 
(a.k.a.\ Slepian--Wolf encoding
\cite[Section 15.8]{CT06}, \cite{SW73}), and so on.
The third message is that this analysis method can be used also
to handle non--trivial 
alternative coding schemes, like a scheme based on the unconstrained
subset sum problem. It turns out that for such a 
scheme, the phase transition
occurs at a critical value of $R$ which is different from the entropy $h(p)$, and we
provide an explicit expression, which is not trivial.

\section{Constrained Subset--Sum Coding}

Consider first the problem of counting the number of solutions $\{\bsigma\}$ to the two
simultaneous equations
\begin{eqnarray}
E(\bsigma)&=&E(\hat{\bsigma})\\
M(\bsigma)&=&M(\hat{\bsigma})
\end{eqnarray}
Denoting the Kronecker Delta function by $\delta(\cdot)$ and $\sqrt{-1}$ by
$i$, we have
\begin{eqnarray}
\Omega(E(\hat{\bsigma}),M(\hat{\bsigma}))&=&
\sum_{\bsigma}\delta(E(\hat{\bsigma})-E(\bsigma))
\delta(
M(\hat{\bsigma})-M(\bsigma))\nonumber\\
&=&\sum_{\bsigma}\delta\left(\sum_ja_j(\hat{\sigma}_j-\sigma_j)\right)
\delta\left(
N_+-N_--\sum_j\sigma_j\right)\nonumber\\
&=&\sum_{\bsigma}\int_{-\pi}^{+\pi}\int_{-\pi}^{+\pi}\frac{\mbox{d}\omega\mbox{d}\theta}{(2\pi)^2}
\exp\left\{i\omega\left[\sum_ja_j(\hat{\sigma}_j-\sigma_j)\right]\right\}\times\nonumber\\
& &\exp\left\{i\theta\left(N_+-N_--\sum_j\sigma_j)\right)\right\}\nonumber\\
&=&\int_{-\pi}^{+\pi}\int_{-\pi}^{+\pi}\frac{\mbox{d}\omega\mbox{d}\theta}
{(2\pi)^2}e^{i\theta(N_+-N_-)}\times\nonumber\\
& &\prod_{j=1}^N\sum_{\sigma_j=-1}^{+1}\exp\{i\omega
a_j(\hat{\sigma}_j-\sigma_j)-i\theta\sigma_j\}\nonumber\\
&=&\int_{-\pi}^{+\pi}\int_{-\pi}^{+\pi}\frac{\mbox{d}\omega\mbox{d}\theta}{(2\pi)^2}e^{i\theta(N_+-N_-)}
\prod_{j=1}^N\left[e^{-i\theta\hat{\sigma}_j}+e^{i(\theta+2\omega
a_j)\hat{\sigma}_j}\right]\nonumber\\
&=&\int_{-\pi}^{+\pi}\int_{-\pi}^{+\pi}
\frac{\mbox{d}\omega\mbox{d}\theta}{(2\pi)^2}e^{i\theta(N_+-N_-)}
\prod_{j:~\hat{\sigma}_j=+1}\left[e^{-i\theta}+e^{i(\theta+2\omega
a_j)}\right]\times\nonumber\\
& &\prod_{j:~\hat{\sigma}_j=-1}\left[e^{i\theta}+e^{-i(\theta+2\omega
a_j)}\right]\nonumber\\
&=&\int_{-\pi}^{+\pi}\int_{-\pi}^{+\pi}
\frac{\mbox{d}\omega\mbox{d}\theta}{(2\pi)^2}\cdot
\prod_{j:~\hat{\sigma}_j=+1}\left[1+e^{2i(\theta+\omega
a_j)}\right]\times\nonumber\\
& &\prod_{j:~\hat{\sigma}_j=-1}\left[1+e^{-2i(\theta+\omega
a_j)}\right].
\end{eqnarray}
Taking now the expectation over $\{a_i\}$, and denoting
\begin{equation}
\phi(\omega)=\left<e^{2i\omega a_1}\right>=\frac{1}{L}\sum_{j=1}^L e^{2i\omega
j},
\end{equation}
we have:
\begin{eqnarray}
\left<\Omega(E(\hat{\bsigma}),M(\hat{\bsigma}))\right>&=&
\int_{-\pi}^{+\pi}\int_{-\pi}^{+\pi}
\frac{\mbox{d}\omega\mbox{d}\theta}{(2\pi)^2}
\left[1+e^{2i\theta}\phi(\omega)\right]^{N_+}\times\nonumber\\
& &\left[1+e^{-2i\theta}\phi(-\omega)\right]^{N_-}\nonumber\\
&=&1+\sum_n\sum_k\left(\begin{array}{cc} N_+ \\ n\end{array}\right)
\left(\begin{array}{cc} N_- \\ k\end{array}\right)\times\nonumber\\
& &\int_{-\pi}^{+\pi}\frac{\mbox{d}\omega}{2\pi}\phi^n(\omega)\phi^k(-\omega)
\int_{-\pi}^{+\pi}\frac{\mbox{d}\theta}{2\pi}e^{2i\theta(n-k)}\nonumber\\
&=&1+\sum_{n=1}^{\min\{N_+,N_-\}}\left(\begin{array}{cc} N_+ \\ n\end{array}\right)
\left(\begin{array}{cc} N_- \\ n\end{array}\right)\times\nonumber\\
& &\int_{-\pi}^{+\pi}\frac{\mbox{d}\omega}{2\pi}[\phi(\omega)\phi(-\omega)]^n\nonumber\\
&=&1+\sum_{n=1}^{\min\{N_+,N_-\}}L^{-2n}\left(\begin{array}{cc} N_+ \\ n\end{array}\right)
\left(\begin{array}{cc} N_- \\
n\end{array}\right)\sum_{s=n}^{nL}(\Lambda_s^n)^2,
\label{longchain1}
\end{eqnarray}
where the summation over $n$ and $k$ in the second line is over 
$\{0,1,\ldots,N_+\}\times
\{0,1,\ldots,N_-\}\setminus\{0,0\}$ and where
$\Lambda_s^n$ is the number of vectors
$\balpha=(\alpha_1,\ldots,\alpha_n)\in\{1,2,\ldots,L\}^n$ 
with $\sum_{i=1}^n\alpha_i=s$.
The last line of eq.\ (\ref{longchain1}) has a simple interpretation:
Let $\calN_+=\{i:~\sigma_i=+1\}$,
$\calN_-=\{i:~\sigma_i=-1\}$,
$\hat{\calN}_+=\{i:~\hat{\sigma}_i=+1\}$, and
$\hat{\calN}_-=\{i:~\hat{\sigma}_i=-1\}$. Obviously,
$E(\bsigma)=E(\hat{\bsigma})$ if and only if
\begin{equation}
\label{event}
\sum_{i\in\calN_+\cap\hat{\calN}_-}a_i=
\sum_{i\in\calN_-\cap\hat{\calN}_+}a_i. 
\end{equation}
Also, for every $\bsigma$ with the
same composition as $\hat{\bsigma}$,
$|\calN_+\cap\hat{\calN}_-|=|\calN_-\cap\hat{\calN}_+|$. 
Let then $n\eqd |\calN_+\cap\hat{\calN}_-|=|\calN_-\cap\hat{\calN}_+|$. 
Every $\bsigma$ with the same composition as 
$\hat{\bsigma}$ corresponds to a choice
of particular subsets ($\calN_+\cap\hat{\calN}_-$ and
$\calN_-\cap\hat{\calN}_+$, both of size $n$) 
of $\hat{\calN}_-$ and $\hat{\calN}_+$,
respectively. For a given $n$, the number of combinations of these subsets is
the product of the binomial coefficients in the last line of
(\ref{longchain1}). For each such combination, the probability of the event
(\ref{event}) is $\sum_s(\Lambda_s^n/L^n)^2$. The last line of eq.\
(\ref{longchain1}) exhausts the product of this probability by the number of
combinations for all possible values of $n$.

So far our analysis has been exact.
We now need an evaluation of the exponential order of $\Lambda_s^n$, where
$s$ scales like $nL$, i.e., $s=\zeta nL$, $\zeta\in (0,1)$, and we expect the
behavior to be symmetric in $\zeta$ about the point $\zeta=1/2$. So it is
enough to cover the range $\zeta\in(0,1/2)$. 
The event $\sum_{i=1}^n\alpha_i=s$ is
obviously equivalent to the event $\sum_{i=1}^n x_i= \zeta n$, where
$x_i=\alpha_i/L$.
The number of points
$\bx=(x_1,\ldots,x_n)\in\{1/L,2/L,\ldots,1-1/L,1\}^n$
with $\sum_{i=1}^nx_i=\zeta n$ is exactly the same as number of points
$(x_1,\ldots,x_{n-1})\in\{1/L,2/L,
\ldots,1-1/L,1\}^{n-1}$ which satisfy $n\zeta-1\le
\sum_{i=1}^{n-1}x_i\le n\zeta$, which is with excellent approximation 
given by
$L^{n-1}$ times the volume of 
the region within the unit cube $[0,1]^{n-1}$ of
continuous valued $(n-1)$--vectors $(y_1,\ldots,y_{n-1})$ that satisfy
$n\zeta-1\le
\sum_{i=1}^{n-1}y_i\le n\zeta$ (see Appendix A for more details).
This volume in turn has an exact formula
(see, e.g., \cite[Theorems 1,4]{MM08}), which is given in the form of a sum of
expressions with alternating signs, but it is not trivial to assess the
exponential order of this formula in a compact manner. 

Alternatively, we may
think of the volume of the set
$\{n\zeta-1\le
\sum_{i=1}^{n-1}y_i\le n\zeta\}$
as the probability
of the event $\{n\zeta-1\le
\sum_{i=1}^{n-1}Y_i\le n\zeta\}$, where $\{Y_i\}$ are i.i.d.\ random variables
all uniformly distributed over $[0,1]$. 
Now, denoting the unit step function
by $u(x)$ and using the fact that it is the inverse Laplace transform of the
function $1/s$, i.e.,
$$u(x)=\frac{1}{2\pi
i}\int_{c-i\infty}^{c+i\infty}\mbox{d}s\cdot\frac{e^{sx}}{s},~~~~~c > 0,$$
we can represent this probability as the following integral in the complex
plane:
\begin{eqnarray}
\label{saddlepoint}
& &\mbox{Pr}\left\{n\zeta-1\le \sum_{i=1}^{n-1}Y_i\le
n\zeta\right\}\nonumber\\
&=&\left<u\left(n\zeta-\sum_{i=1}^{n-1}Y_i\right)-
u\left(n\zeta-1-\sum_{i=1}^{n-1}Y_i\right)\right>\nonumber\\
&=&\left<\frac{1}{2\pi i}
\int_{c-i\infty}^{c+i\infty}\mbox{d}s\exp\left[s\left(n\zeta-\sum_{i=1}^{n-1}Y_i\right)
\right]\cdot\frac{(1-e^{-s})}{s}\right>\nonumber\\
&=&\frac{1}{2\pi i}
\int_{c-i\infty}^{c+i\infty}\mbox{d}se^{s\zeta
n}\left<\exp\left(-s\sum_{i=1}^{n-1}Y_i\right)
\right>\cdot\frac{(1-e^{-s})}{s}\nonumber\\
&=&\frac{1}{2\pi i}\int_{c-i\infty}^{c+i\infty}\mbox{d}se^{s\zeta
n}\left<
e^{-sY_1}\right>^{n-1}
\cdot\frac{(1-e^{-s})}{s}\nonumber\\
&=&\frac{1}{2\pi i}\int_{c-i\infty}^{c+i\infty}\mbox{d}se^{s\zeta
n}\left(\frac{1-e^{-s}}{s}
\right)^{n-1}
\cdot\frac{(1-e^{-s})}{s}\nonumber\\
&=&\frac{1}{2\pi i}\int_{c-i\infty}^{c+i\infty}\mbox{d}se^{s\zeta
n}\left(\frac{1-e^{-s}}{s}
\right)^n\nonumber\\
&=&\frac{1}{2\pi i}\int_{c-i\infty}^{c+i\infty}
\mbox{d}s\exp\left\{n\left[s\zeta+\ln(1-e^{-s})-\ln
s\right]\right\}.
\end{eqnarray}
This integral is easily evaluated using the saddle point method. 
In Appendix B, we show that the highest modulus of the integrand
along the integration path, which is the
vertical straight line $\mbox{Re}(s)=c$, is uniquely obtained
at $s=c$,
which yields
\begin{equation}
\mbox{Pr}\left\{n\zeta-1\le \sum_{i=1}^{n-1}Y_i\le n\zeta\right\}\exe
e^{-n\Phi(\zeta)}
\end{equation}
where
\begin{equation}
\Phi(\zeta)=\max_{t\ge 0}[\ln t-\ln(1-e^{-t})-\zeta t], ~~~\zeta\in(0,1/2)
\end{equation}
and we extend the definition of $\Phi(\cdot)$ to the interval $(0,1)$
to be symmetric around $\zeta=1/2$, namely, $\Phi(1/2-\zeta)=\Phi(\zeta)$.
Note that maximizing $t$ is the saddle point of the integral
(\ref{saddlepoint}), namely, it
is the point at which the derivative of the expression
in the square brackets vanishes. Also, the axis 
\cite[Section 5.4, p.\ 84]{deBruijn81} 
of this saddle point is in the vertical direction, which is
the natural direction of integration path anyway.
Thus, we now have
\begin{equation}
\Lambda_s^n\approx L^{n-1}\cdot e^{-n\Phi(\zeta)}; ~~~~~~s=\zeta
nL,~~\zeta\in(0,1).
\end{equation}
On substituting this into the inner summation of
(\ref{longchain1}), we get
\begin{eqnarray}
\sum_{s=n}^{nL}(\Lambda_s^n)^2&\exe&
L^{2(n-1)}\sum_{s=n}^{nL}
\exp\left\{-n\Phi\left(\frac{s}{nL}\right)
\right\}\nonumber\\
&=&nL^{2n-1}\sum_{s=n}^{nL}\frac{1}{nL}
\exp\left\{-n\Phi\left(\frac{s}{nL}\right)\right\}\nonumber\\
&\exe&L^{2n-1}\int_0^1\mbox{d}\zeta
\exp\{-n\Phi(\zeta)\}\nonumber\\
&\exe& L^{2n-1}e^{-n\inf_{0< \zeta< 1}\Phi(\zeta)}\nonumber\\
&\exe& L^{2n-1},
\end{eqnarray}
where the last step follows from the fact that the infimum of $\Phi(\zeta)$
over $\zeta\in(0,1)$ is zero (achieved at $\zeta=1/2$).
Thus,
\begin{eqnarray}
\left<\Omega(E(\hat{\bsigma}),M(\hat{\bsigma}))\right>&\doteq&1+
\frac{1}{L}\sum_n\left(\begin{array}{cc} N_+ \\ n\end{array}\right)
\left(\begin{array}{cc} N_- \\
n\end{array}\right)\nonumber\\
&=&1+\frac{1}{L}\cdot\exp_2\left\{N\sup_{0<\alpha<\min\{p,q\}}\left[ph\left(\frac{\alpha}{p}\right)+
qh\left(\frac{\alpha}{q}\right)\right]\right\}\nonumber\\
&=&1+\frac{1}{L}\cdot 2^{Nh(p)}.
\end{eqnarray}
Thus, for $L=2^{NR}$, we have
\begin{equation}
\left<\Omega(E(\hat{\bsigma}),M(\hat{\bsigma}))\right>-1\exe 2^{N[h(p)-R]},
\end{equation}
which means that there is a phase transition at the critical rate of
$R_c=h(p)$, as mentioned earlier.
For $R > h(p)$, the expression $\left<\Omega(E(\hat{\bsigma}),M(\hat{\bsigma}))\right>-1$
decays exponentially,
and hence so does 
$\mbox{Pr}\{\Omega(E(\hat{\bsigma}),M(\hat{\bsigma})) > 1\}$,
which means that the decoding is unambiguous and correct
with high probability. On the other hand, for $R < h(p)$ the probability for
the existence of many additional solutions $\{\bsigma\}$ must be very
high: Since the number of typical source sequences
(i.e., sequences with $M(\bsigma)=M(\hat{\bsigma})=N(p-q)$) is exponentially
$2^{Nh(p)}$ and the number of distinct values of $E(\bsigma)$ cannot
exceed $N\cdot L =N\cdot 2^{NR}$, the fraction of sequences
$\{\bsigma\}$ (with the given composition) that
are {\it unique} solutions to the equation 
$E(\bsigma)=E(\hat{\bsigma})$ cannot
exceed $N2^{NR}/2^{Nh(p)}\exe 2^{-N[h(p)-R]}$, and so,
$\mbox{Pr}\{\Omega(E(\hat{\bsigma}),M(\hat{\bsigma})) > 1\}\ge
1-2^{-N[h(p)-R]}$.

This result is actually quite expected. The critical rate $R_c$ cannot be
strictly larger than $h(p)$ 
because, as said, the total number of sequences with composition $(p,q)$
is exponentially $2^{Nh(p)}$: Had $R_c$ been larger than $h(p)$ we would
have obtained that $\left<\Omega(\hat{E})\right>$ grows with an exponential
rate which is faster than $h(p)$ (at least when $L$ is subexponential),
which is impossible.
On the other hand, $R_c$ cannot be strictly smaller than $h(p)$, because
then it would mean that Sasmoto's coding scheme achieves a compression
ratio that is better than the entropy.
Thus, $R_c$ must be equal to $h(p)$.

The above derivation extends straightforwardly in several directions
(one at a time or simultaneously):\\

\vspace{0.2cm}

\noindent
{\it 1. Lossless source coding in
the presence of correlated side information at the decoder.} 
Consider the case where the decoder has access to a side information sequence
$\btau=(\tau_1,\ldots,\tau_N)$, which is correlated to the source sequence
according to a given joint distribution $P(\sigma,\tau)$,
and the two sequences are i.i.d.\ over time, i.e.,
\begin{equation}
P(\bsigma,\btau)=\prod_{i=1}^N P(\sigma_i,\tau_i).
\end{equation}
It is well known (see, e.g., \cite[Section 15.8]{CT06})
that in this case, the best achievable compression ratio
is given by the conditional entropy of the source
given the side information, even if the encoder does not have access to
the side information (Slepian--Wolf coding \cite{SW73}). 
Here we propose an alternative
way to achieve this optimum compression ratio based on subset sum encoding:
The encoder works
essentially in the same manner as before. The decoder seeks solutions to the
equation $\sum_ia_i\sigma_i=E(\hat{\bsigma})$ 
only within the set of $\bsigma$--vectors whose joint empirical distribution
together with the side information sequence is 
close to the joint distribution 
$P(\sigma,\tau)$ (within some small tolerance). In this case, 
an analysis similar to the above, reveals that the critical
rate is given by the conditional entropy of the source given the side
information.\\

\vspace{0.2cm}

\noindent
{\it 2. Multiple Subset Sums.} Instead of one set of random variables
$a_1,\ldots,a_N$, randomly drawn in $\{1,2,\ldots,L\}$, consider an array of
random variables $\{a_i^k\}$, $i=1,2,\ldots,N$, $k=1,2,\ldots,m$, all
statistically independent, where each $a_i^k$ is uniformly distributed
in $\{1,2,\ldots,L_k\}$, where $L_k=2^{NR_k}$, $R_k > 0$. The encoder encodes
$\bsigma$ into $(M(\bsigma),
E_1(\bsigma),\ldots,E_m(\bsigma))$, where each
$E_k(\sigma)\eqd\sum_ia_i^k\sigma_i$ is represented by
$\log (NL_k)=\log N+NR_k$ bits. The decoder reconstructs $\bsigma$ as the
first vector whose encoding agrees with the given compressed input
$(M(\bsigma),
E_1(\bsigma),\ldots,E_m(\bsigma))$. 
It is easy to show that the decoding is unambiguous with high probability iff
$\sum_k R_k > H$. Thus, here the phase transition occurs 
at the whole hyperplane $\sum_kR_k=H$. 
%This structure has the advantage of
%reducing the search complexity at the encoder. For example, consider the
%case $m=2$. The decoder first makes a list of all sequences with
%$M(\bsigma)=M(\hat{\bsigma})$ for which $E_1(\bsigma)$ agrees with that of the
%compressed representation. This takes

\vspace{0.2cm}

\noindent
{\it 3. General finite alphabets.}
Another natural direction of extending the above result is from the case of a
binary source alphabet to a general finite alphabet
which, without loss of generality, will be assumed to be
$\Sigma=\{1,2,\ldots,K\}$.
A simple strategy is to decompose the problem into $K-1$ binary encoding
problems, and in each one of them we can
rely directly on the binary code construction.
Let the input source string $\bsigma$ have $N_s=Np_s$ occurrences of
$\sigma_i=s\in\Sigma$, $s=1,2,\ldots,K$ (of course, $\sum_{s=1}^KN_s=N$).
Now, for each $s=1,2,\ldots,K-1$, let us select
independently at random $a_1^s,a_2^s,\ldots,a_N^s$, each one drawn under the uniform
distribution over the integers $\{1,2,\ldots,L_s\}$, where $L_s=2^{NR_s}$ and
$R_s$ will be
specified shortly. The resulting random selections are revealed to the encoder
and the decoder. The encoder works as follows: Given the source vector
$\bsigma$, we first encode
the numbers $N_1,N_2,\ldots,N_{K-1}$ 
as overhead (just like the transmission of $M(\sigma)$ in
the binary case). Next, for each $s=1,\ldots,K-1$, we calculate
$E_s(\bsigma)=\sum_{i:~\sigma_i=s}a_i^s-
\sum_{i:~\sigma_i> s}a_i^s$ and transmit
its value using $\log(N\cdot L_s)$ bits. The role of each $E_s(\bsigma)$ is to represent the
information pertaining to all locations where $\sigma_i=s$. Based on
the results of the binary alphabet case, in order to
decode $E_1(\bsigma)$ unambiguously, $R_1$ 
should be at least as large as $h(p_1)$
(think of encoding the binary sequences $\{\calI(\sigma_i=1)\}$,
where $\calI(\cdot)$ is the indicator function). 
For the next stage, the task is to fill in $N_2$ out of the remaining $(N-N_1)$ locations
by $s=2$, and so we have reduced the problem to that of
encoding the binary sequence $\{\calI(\sigma_i=2)\}_{i:~\sigma_i\ne 1}$ of length
$(N-N_1)$. By the same reasoning then,
to decode $E_2(\bsigma)$ reliably, $R_2$ should be at least as large as $(1-p_1)h(p_2/(1-p_1))$.
This procedure continues until $s=K-1$, where reliable decoding of
$E_{K-1}(s)$ requires
$R_{K-1}>(1-p_1-\ldots-p_{K-2})h(p_{K-1}/(1-p_1-\ldots-p_{K-2}))$.
The overall coding rate (neglecting the overhead) is then
$$h(p_1)+(1-p_1)h\left(\frac{p_2}{1-p_1}\right)+\ldots+
(1-p_1-\ldots-p_{K-2})h\left(\frac{p_{K-1}}{1-p_1-\ldots-p_{K-2}}\right),$$
which is easily shown (using the chain rule of the entropy) 
to be identical to the entropy of the source
$$H=-\sum_{s=1}^Kp_s\log p_s.$$

\section{Unconstrained Subset--Sum Coding}

Returning to the binary case,
suppose next that we wish to make the mapping from $\bsigma$ to $E(\bsigma)$ essentially
one--to--one over the {\it entire} source
vector space $\{-1,+1\}^N$, and then there would be no need to specify the composition of
$\hat{\bsigma}$ to the decoder. This corresponds to the unconstrained subset
sum problem. How large should $R$ be now? Here, the analysis is similar but somewhat more
involved. The point in presenting the analysis for this case is not quite
motivated by the
usefulness of the data
compression scheme itself, but more about demonstrating the applicability of
the analysis method. This time, the derivation is as follows:
\begin{eqnarray}
\Omega(E(\hat{\bsigma}))
&=&\sum_{\bsigma}\delta\left(\sum_{j=1}^N
a_j(\hat{\sigma}_j-\sigma_j)\right)\nonumber\\
&=&\sum_{\bsigma}\int_{-\pi}^{+\pi}\frac{\mbox{d}\omega}{2\pi}\cdot
e^{i\omega\sum_{j=1}^Na_j(\hsig_j-\sigma_i)}\nonumber\\
&=&\int_{-\pi}^{+\pi}\frac{\mbox{d}\omega}{2\pi}\cdot\sum_{\bsigma}
\prod_{j=1}^Ne^{i\omega a_j(\hsig_j-\sigma_i)}\nonumber\\
&=&\int_{-\pi}^{+\pi}\frac{\mbox{d}\omega}{2\pi}
\prod_{j=1}^N\left[\sum_{\sigma_j}e^{i\omega a_j(\hsig_j-\sigma_i)}\right]\nonumber\\
&=&\int_{-\pi}^{+\pi}\frac{\mbox{d}\omega}{2\pi}
\prod_{j=1}^N\left(1+e^{i2\omega a_j\hsig_j}\right).
\end{eqnarray}
Taking now the expectation w.r.t.\ the randomness of $\{a_j\}$,
we readily obtain
\begin{equation}
\left<\Omega(E(\hat{\bsigma}))\right>_{\ba}
=\int_{-\pi}^{+\pi}\frac{\mbox{d}\omega}{2\pi}
\left[1+\phi(\omega)\right]^{N_+}\cdot\left[1+\phi^*(\omega)\right]^{N_-}.
\end{equation}
Assuming that the binary vector $(\hsig_1,\ldots,\hsig_N)$ is governed by a
binary memoryless source with
$p=\mbox{Pr}\{\hsig_i=+1\}=1-\mbox{Pr}\{\hsig_i=-1\}=1-q$, we next take an
ensemble average w.r.t.\ the randomness of $\{\hsig_i\}$, and obtain
\begin{eqnarray}
\label{longchain2}
\left<\Omega(E(\hat{\bsigma}))\right>&=&
\int_{-\pi}^{+\pi}\frac{\mbox{d}\omega}{2\pi}
\sum_{k=0}^N\left(\begin{array}{cc}N \\
k\end{array}\right)\left[p(1+\phi(\omega))\right]^k
\cdot\left[q(1+\phi^*(\omega))\right]^{N-k}\nonumber\\
&=&\int_{-\pi}^{+\pi}\frac{\mbox{d}\omega}{2\pi}
\left[1+p\phi(\omega)+q\phi^*(\omega)\right]^N\nonumber\\
&=&1+\sum_{k=1}^N\left(\begin{array}{cc}N \\ k\end{array}\right)\cdot
\int_{-\pi}^{+\pi}\frac{\mbox{d}\omega}{2\pi}\left[p\phi(\omega)+q\phi^*(\omega)\right]^k\nonumber\\
&=&1+\sum_{k=1}^N\left(\begin{array}{cc}N \\ k\end{array}\right)\cdot
L^{-k}\int_{-\pi}^{+\pi}\frac{\mbox{d}\omega}{2\pi}
\left[p\sum_{\ell=1}^{L}e^{i2\omega\ell}+q\sum_{\ell=1}^{L}e^{-i2\omega\ell}\right]^k\nonumber\\
&=&1+\sum_{k=1}^N\left(\begin{array}{cc}N \\ k\end{array}\right)\cdot L^{-k}
\sum_{r=0}^k\left(\begin{array}{cc}k \\
r\end{array}\right)p^rq^{k-r}\times\nonumber\\
& &\int_{-\pi}^{+\pi}\frac{\mbox{d}\omega}{2\pi}
\left(\sum_{\ell=1}^{L}e^{i2\omega\ell}\right)^r\cdot
\left(\sum_{\ell=1}^{L}e^{-i2\omega\ell}\right)^{k-r}\nonumber\\
&=&1+\sum_{k=1}^N\left(\begin{array}{cc}N \\ k\end{array}\right)\cdot L^{-k}
\sum_{r=0}^k\left(\begin{array}{cc}k \\
r\end{array}\right)p^rq^{k-r}\times\nonumber\\
& &\int_{-\pi}^{+\pi}\frac{\mbox{d}\omega}{2\pi}
\left(\sum_{\ell=r}^{rL}\Lambda_{\ell}^re^{i2\omega\ell}\right)\cdot
\left(\sum_{\ell=k-r}^{(k-r)L}\Lambda_{\ell}^{k-r}e^{-i2\omega\ell}\right)\nonumber\\
&=&1+\sum_{k=1}^N\left(\begin{array}{cc}N \\ k\end{array}\right)\cdot L^{-k}
\sum_{r=0}^k\left(\begin{array}{cc}k \\
r\end{array}\right)p^rq^{k-r}\times\nonumber\\
& &\sum_{\ell=\max\{r,k-r\}}^{L\min\{r,k-r\}}
\Lambda_{\ell}^r\Lambda_{\ell}^{k-r}.
\end{eqnarray}
Now, similarly as before
\begin{eqnarray}
& &\sum_{\ell=\max\{r,k-r\}}^{L\min\{r,k-r\}}\Lambda_{\ell}^r\Lambda_{\ell}^{k-r}\nonumber\\
&\exe&
L^{k-2}\sum_{\ell=\max\{r,k-r\}}^{L\min\{r,k-r\}}
\exp\left\{-r\Phi\left(\frac{\ell}{rL}\right)-(k-r)
\Phi\left(\frac{\ell}{(k-r)L}\right)\right\}\nonumber\\
&=&L^{k-1}\sum_{\ell=\max\{r,k-r\}}^{L\min\{r,k-r\}}\frac{1}{L}
\exp\left\{-r\Phi\left(\frac{\ell}{rL}\right)-(k-r)
\Phi\left(\frac{\ell}{(k-r)L}\right)\right\}\nonumber\\
&\exe&L^{k-1}\int_0^{\min\{r,k-r\}}\mbox{d}x
\exp\left\{-r\Phi\left(\frac{x}{r}\right)-(k-r)
\Phi\left(\frac{x}{k-r}\right)\right\}\nonumber\\
&\exe& L^{k-1}e^{-k\psi(\beta)},
\end{eqnarray}
where 
$$\psi(\beta)=\min_{0\le
x\le\min\{\beta,1-\beta\}}\left[\beta\Phi\left(\frac{x}{\beta}\right)+
(1-\beta)\Phi\left(\frac{x}{1-\beta}\right)\right],$$
where $\beta\eqd r/k$. Denoting $\alpha=k/N$ and
substituting this into eq.\ (\ref{longchain2}), we obtain
\begin{eqnarray}
& &\left<\Omega(E(\hat{\bsigma}))\right>-1\nonumber\\
&\exe&\frac{1}{L}\cdot\sum_{k=1}^N
\left(\begin{array}{cc}N \\ k\end{array}\right)\sum_{r=0}^k
\left(\begin{array}{cc}k \\
r\end{array}\right)p^rq^{k-r}e^{-k\psi(\beta)}\nonumber\\
&\exe&\frac{1}{L}\cdot\exp\left\{N\max_\alpha[h(\alpha)+\alpha\max_\beta(h(\beta)+\beta\ln
p+(1-\beta)\ln q-\psi(\beta))]\right\}\nonumber\\
&=&\frac{1}{L}\cdot\exp\left\{N\max_\alpha[h(\alpha)-
\alpha\min_\beta(D(\beta\|p)+\psi(\beta))]\right\}\nonumber\\
&=&\frac{1}{L}\cdot\exp\left\{N\max_\alpha[h(\alpha)-
\alpha\xi(p)]\right\}\nonumber\\
&=&\frac{1}{L}\cdot 2^{N\log_2[1+e^{-\xi(p)}]},
\end{eqnarray}
where we have defined
\begin{equation}
D(\beta\|p)=\beta\ln\frac{\beta}{p}+(1-\beta)\ln\frac{1-\beta}{1-p}
\end{equation}
and
\begin{equation}
\xi(p)=\min_\beta[D(\beta\|p)+\psi(\beta)].
\end{equation}
Again, if $L=2^{NR}$, then the phase transition is now at
$R=R_c$, where
\begin{equation}
R_c=\log_2[1+e^{-\xi(p)}].
\end{equation}

Note the special care should be exercised in the
extreme cases where $p=0$ and $p=1$. It is easy to see that in these
cases $D(\beta\|p)=\infty$ for all $\beta$, except $\beta=p$ but when
$\beta=0$ and $\beta=1$, $\psi(\beta)=\infty$, thus the sum
$D(\beta\|p)+\psi(\beta)$ is infinite for every $\beta\in[0,1]$.
The choice $\alpha=0$ is actually not allowed
in the out--most 
maximization over $\alpha$ since the sum over $k$ begins with $k=1$.
Thus, for $p=0$ and $p=1$, we have $R_c=-\infty$, which means that
$\left<\Omega(\hat{E})\right>=1$, as expected.

Obviously, $R_c$ cannot be smaller than the entropy of the
source.\footnote{
Had it been
smaller, we could have compressed at a rate below the entropy using Sasamoto's
coding scheme, which is a contradiction \cite{CT06}.}
In general, $R_c$ is larger than the entropy, but
the coding rate of the
corresponding data compression scheme
need not be as large as $R_c$.
By applying variable--rate entropy coding to $E(\bsigma)$, taking advantage
of the non--uniform distribution of this random variable, one can compress
it at a rate very close to the entropy of the source. While in this case,
$R_c$ no longer has the meaning of coding rate, it does another meaning,
which is related to the storage requirement for saving the numbers
$\{a_i\}$. Note that
for $R=1$, it is easy to specify a 
particular set of integers $\{a_i\}$ that yields a
one--to--one mapping from $\bsigma$ to $E(\bsigma)$: By setting $a_i=2^{i-1}$,
$E(\bsigma)$ becomes the standard binary representation of $\bsigma$ (up to a
fixed shift). The above result tells us that we can manage with less
storage since $R_c$ is in general less than 1, except the case $p=1/2$, where
$R_c=1$ (see also \cite[Proposition 7.6 and Exercise 7.8]{MM09}).

\section*{Acknowledgment} Useful discussions with Alexander Vardy and with Tuvi
Etzion, concerning the evaluation of $\Lambda_s^n$ 
(which appears first in eq.\ (\ref{longchain1})), are acknowledged with
thanks. In particular, A.~Vardy has drawn my attention to ref.\ \cite{MM08}
in this context and pointed out its relevance.

\section*{Appendix}

\subsection*{A. Estimating $\Lambda_s^n$}

Given a lattice and given a certain region in space, the number of lattice
sites in that region is approximately given by the volume of the region
divided by the volume of a single Voronoi cell pertaining to that lattice.
This approximation improves as the ratio between these two volumes becomes
very large. More precisely,
there is a slight correction due to the fact that some of
these Voronoi cells may not be entirely included in the region in question. To
obtain upper and lower bounds on the number of lattice points, one may
slightly expand (for an upper bound)
or shrink (for a lower bound) the given region by an amount that guarantees
full inclusion (for an upper lower bound) or full exclusion (for a lower
bound)
of the partially included Voronoi
cells that are near the boundaries. In our case, we have a cubic lattice in
$(n-1)$ dimensions with spacings of $1/L$ in each dimension, so the volume of a
Voronoi cell is $1/L^{n-1}$. By replacing $n\zeta$ with $n\zeta \pm n/L$,
one can obtain the aforementioned upper and lower bounds, but the
corrections of $\pm n/L$ to $n\zeta$ and to $n\zeta-1$ have negligible effects
in the exponential
scale since $L$ grows exponentially with $n$
and hence $n/L$ vanishes in the limit.

\subsection*{B. Saddle Point of the Integral (\ref{saddlepoint})}

The modulus of the integrand depends solely on the real part of the exponent
of the integrand, namely, on $\mbox{Re}\{s\zeta+\ln(1-e^{-s})-\ln s\}$. Now, consider an
arbitrary complex number $s=r+i\omega$. Then obviously,
\begin{equation}
\mbox{Re}\{s\zeta+\ln(1-e^{-s})-\ln s\}=r \zeta
+\frac{1}{2}\ln\left(\frac{1+e^{-2r}-2e^{-r}\cos\omega}{r^2+\omega^2}\right).
\end{equation}
Thus, we have to show that
$$\frac{1+e^{-2r}-2e^{-r}\cos\omega}{r^2+\omega^2}$$
is maximized at $\omega=0$, and only at $\omega=0$, for all $r > 0$.
This would be equivalent
to the assertion that
\begin{equation}
\label{needtobeproved}
\frac{(1-e^{-r})^2}{r^2}\ge
\frac{1+e^{-2r}-2e^{-r}\cos\omega}{r^2+\omega^2},
\end{equation}
with equality if and only if $\omega=0$,
since the left--hand side is obtained from the right--hand side by setting
$\omega=0$. 

In order to prove eq.\ (\ref{needtobeproved}),
we begin from the obvious inequality $\theta
\ge \sin(\theta)$, which holds for all $\theta \ge 0$. Integrating both
sides of this inequality from $0$ to $\omega$ ($\omega > 0$), we obtain
$\omega^2/2\ge 1-\cos(\omega)$ with equality if and only 
if $\omega=0$, or equivalently,
\begin{equation}
\label{omegaineq}
\frac{\omega^2}{1-\cos(\omega)}\ge 2,
\end{equation}
which now holds for all $\omega$ since the left--hand side
is an even function. Similarly, beginning from the inequality $1\le\cosh(r)$
and integrating both sides twice, we get
$r^2/2\le \cosh(r)-1$, or equivalently
\begin{equation}
\label{rineq}
\frac{r^2}{\cosh(r)-1}\le 2,
\end{equation}
which when combined with (\ref{omegaineq}) yields
\begin{equation}
\frac{r^2}{\cosh(r)-1}\le \frac{\omega^2}{1-\cos(\omega)},
\end{equation}
or equivalently,
\begin{equation}
\frac{2r^2e^{-r}}{(1-e^{-r})^2}\le \frac{\omega^2}{1-\cos(\omega)},
\end{equation}
which is 
\begin{equation}
2r^2e^{-r}-2r^2e^{-r}\cos(\omega) \le \omega^2(1-e^{-r})^2.
\end{equation}
Adding $r^2(1-e^{-r})^2$ to both sides 
of this inequality
and rearranging terms,
one obtains
\begin{equation}
r^2[1+e^{-2r}-2e^{-r}\cos(\omega)]\le (r^2+\omega^2)(1-e^{-r})^2,
\end{equation}
which is equivalent to eq.\ (\ref{needtobeproved}).

\end{document}